\documentclass[superscriptaddress,showpacs,preprint,pra]{revtex4-1} 
\usepackage{epsfig}
\usepackage{amsfonts}
\usepackage{amsmath}
\usepackage{xcolor}

\begin{document}

\title{Exact wave functions and entropies of the one dimensional Regularized Calogero model}

\author{Federico M. Pont}
\email{pont@famaf.unc.edu.ar}
\affiliation{Facultad de Matem\'atica, Astronom\'{\i}a y F\'{\i}sica,
Universidad Nacional de C\'ordoba and IFEG-CONICET, Ciudad Universitaria,
X5016LAE C\'ordoba, Argentina}

\author{Omar Osenda}
\email{osenda@famaf.unc.edu.ar}
\affiliation{Facultad de Matem\'atica, Astronom\'{\i}a y F\'{\i}sica,
Universidad Nacional de C\'ordoba and IFEG-CONICET, Ciudad Universitaria,
X5016LAE C\'ordoba, Argentina}

\author{Pablo Serra}
\email{serra@famaf.unc.edu.ar}
\affiliation{Facultad de Matem\'atica, Astronom\'{\i}a y F\'{\i}sica,
Universidad Nacional de C\'ordoba and IFEG-CONICET, Ciudad Universitaria,
X5016LAE C\'ordoba, Argentina}

\begin{abstract}
The divergence in the interaction term of the Calogero model can be prevented 
introducing a cutoff length parameter, this modification leads to a 
quasi-exactly solvable model whose eigenfunctions can be written in terms of 
Heun's polynomials. It is shown both, analytical and numerically. that the 
reduced density matrix obtained tracing out one particle from the two-particle 
density operator can be obtained exactly as well as its entanglement spectrum. 
The number of non-zero eigenvalues in these cases is finite. Besides, it is 
shown that taking the limit in which the cutoff distance goes to zero, the 
reduced density matrix and finite entanglement spectrum of the Calogero model 
is retrieved. The entanglement R\'enyi entropy is also studied to characterize the physical 
traits of the model. It is found that the quasi-exactly solvable character of 
the model is put into evidence by the entanglement entropies when they are 
calculated numerically over the parameter space of the model. 
\end{abstract}
\date{\today}

\maketitle

\section{Introduction}

The Calogero model \cite{Calogero1969} occupies a remarkable place in 
theoretical and 
mathematical physics. It has been linked, or is an integral part of 
advances made in quantum Hall effect \cite{Azuma1994}, random matrices 
\cite{Altshuler1993}, integrability \cite{Poly1993}, Yang-Mills theory 
\cite{Gorsky1994}, etc.

Remarkably, the Calogero model variants, the deformed  \cite{Chalykh1998, 
Atai2017}, 
the different generalizations \cite{Polychronakos1999,Znojil2001} and the 
rectified ones \cite{Downing2017}, inherit many of 
its properties, a trend that was acknowledged from very early by Sutherland 
\cite{Sutherland1971}. More recently, it has been shown that the  
$p-$reduced density matrix (p-RDM) of a $N$-particle one-dimensional Calogero 
model can also be obtained exactly \cite{opos15}, as well as the entanglement 
spectrum, for a discrete set of the strength interaction parameter (let us 
remember that the p-RDM matrix is obtained when $(N-p)$ particles are traced 
out of 
the density matrix of an $N-$particle system). Besides, at 
these values the R\'enyi entanglement entropies show non-analytical behaviour 
\cite{Garagiola2016} in contradistinction with the von Neumann entropy.That the Calogero model admits 
exact reduced density matrices was also noted by Katsura and Hatsuda 
\cite{Katsura2007}.

The harmonic confinement potential, present in all the variants of the model, 
was included more as a mean to keep the particles bounded, since the 
interaction between them is mainly repulsive, rather than as a model for an 
implementable potential. Nevertheless, the advances made in cold confined gases 
and quantum dots make the model applicable to analyze actual experimental 
setups.

Quasi-exactly solvable models \cite{quasi1}, as the so called spherium 
\cite{Ezra1982,Ezra1983,Loos2009}, or other electron systems confined in 
boxes with different geometries, as square \cite{Alavi2000,Ghosh2005}, cylindrical 
\cite{Ryabinkin2010} and spherical \cite{Jung2004} are of interest because they 
are more amenable of analytical treatment than boundless ones. Besides its 
behaviour is quite 
different from the observed in extended systems, so they posed new challenges 
that allow to improve the DFT method \cite{Loos2014a,Loos2014b}. 
They are benchmarks where many numerical 
methods can be tested 
\cite{Loos2010,Jung2004,Jung2003} and model strongly confined electron systems. 
The condition 
of quasi-exactly solvable means that the spectrum and the eigenfunction are 
exactly known in a discrete set of the Hamiltonian parameters. Recently there 
has been a flurry 
of activity in this subject 
\cite{Loos2009,Downing2016,Tognetti2016,Downing2013,Downing2017}, while early 
examples, 
dealing with the same quasi-exactly solvable model, can be found in  the works 
of Kais \emph{et. al.} \cite{Kais1989} and Taut \cite{Taut1993}. The recent advances made in 
one- and 
two-particle quasi-exactly solvable models rely heavily on the properties of 
the polynomial solutions of the Heun differential equation \cite{Heun1}. 

The broad application to many different problems in classical and quantum 
physics of 
the Heun'equations and its polynomial solutions has been made possible by the 
work of Fiziev~\cite{Fiziev2010}, in particular to the dynamics of a rotor 
vibratory giroscope~\cite{Motsepe2014}, the calculation of natural occupation 
numbers in two electron quantum-rings~\cite{Tognetti2016}, the solution of the 
Schr\"odinger equation for a particle trapped in a hyperbolic double-well 
potential~\cite{Downing2013}, the problem of two electrons confined on a 
hypersphere~\cite{Loos2009}, one electron in crossed inhomogeneous magnetic and 
homogeneous electric fields~\cite{Downing2016}, and in the study of normal modes 
in non-rotating black holes~\cite{Fiziev2011}. The recent and salient role 
played 
by the Heun functions, and its foreseeable future, in natural sciences is 
depicted in the Introduction of Ref. \cite{Fiziev2012}.

Quite recently, Downing has reported the analytical solutions of a two-electron 
quantum model \cite{Downing2017}. In the model analyzed, the two particles 
interact via a regularized potential that decays as the inverse of the distance 
between the particles squared, {\em i.e.} the model is a regularized 
Calogero model. The regularization prevents the divergence of the potential 
when the distance between the particles goes to zero, introducing a short 
distance cutoff parameter $d$. Remarkably, the model is quasi-exactly solvable 
so, for a given value of $d$, the exact two-particle wave function can be 
obtained only for a discrete set of values of the interaction strength 
parameter $g$, as is usually denoted in the context of the Calogero model. 
At these values, the two-particle wave function is a polynomial function of 
the inter-particle distance. As has been shown in Ref.~\cite{opos15}, when 
the wave function of a multi-particle Calogero model can be written as the 
product of  a polynomial function depending on the inter-particles distances, 
times other functions that depend separately on the coordinates of each 
particle, then the $p$-RDM and the entanglement spectrum can be both obtained 
exactly. Since  the R\'enyi entropy also shows a very particular  behavior 
for these discrete set of values, it does beg the question of how many of these 
features are inherited by the regularized model. To this end, we study the 
exact solutions of the one-dimensional two particle regularized model, both
their symmetric and anti-symmetric solutions under particle interchange and 
their reduced density matrices. Even though this is a quasi-exactly solvable model we calculate numerical solutions
to study the whole Hamiltonian parameter space. A motivation to perform such exact calculations of
entropies arises from the need of stringent benchmarks that assess the accuracy of numerical approximations
when they are applied to confined electron systems~\cite{Jiao2017}.

This paper is organized as follows: The regularized Calogero model is presented in 
Section~\ref{sec:model}. In Section~\ref{sec:symmetric} the exact symmetric wave functions are thoroughly 
analyzed while the antisymmetric ones are the subject of Section~\ref{sec:anti-symmetric}. The 
R\'enyi and von Neumann entropies for the exact two-particle states, together with numerical approximations,
 are presented in Section~\ref{sec:renyi}. Finally, a discussion of the results and some open questions are presented in
 Section~\ref{sec:discussion}.

\section{The model and its eigenfunctions}\label{sec:model}

Recently, Downing \cite{Downing2017} showed that the three-dimensional two-particle regularized 
Calogero model is solvable for a discrete set of values of the interacting parameter.
In this work, we address the one dimensional two-particle regularized Calogero 
Hamiltonian 
\begin{equation}
\label{eq:ehcal}
H = h(1)\,+\, h(2) +  \frac{g}{x_{12}^2+2 \,d^2} \,,
\end{equation}
 where
\begin{equation}
\label{ehi}
h(i) = \frac{1}{2}\, p_i^2 + \frac{1}{2}\,x_i^2\;\;\mbox{and}\;\;\,x_{12}\,=\,|x_1-x_2|\, .
\end{equation}
\noindent In particular, we look for a discrete set of exact two-particle symmetric or 
antisymmetric wave functions. We do not assume particular values for the spin 
variable, so the symmetric and antisymmetric functions can be used to 
construct two-fermions or two-bosons solutions depending on the symmetry of the 
spinorial part of the quantum state.

With the coordinate transformation

\begin{equation}
X\,=\,\frac{1}{\sqrt{2}} (x_1+x_2) \;\;\;;\;\;\;x\,=\,\frac{1}{\sqrt{2}} (x_1-x_2)\,,
\end{equation} 

\noindent the Hamiltonian Eq.~(\ref{eq:ehcal}) takes the form $H=H_X+H_x$, where

\begin{subequations}
\begin{eqnarray}
H_X\,=\,-\frac{1}{2} \frac{d^2}{dX^2} +\frac{1}{2} X^2 \;\;\;; \label{ehX}\\
H_x\,=\,-\frac{1}{2} \frac{d^2}{dx^2} +\frac{1}{2} x^2 + \frac{g/2}{x^2+d^2} \label{ehx}\,
.
\end{eqnarray}
\end{subequations}

The eigenfunctions will be the product of eigenfunctions of each Hamiltonian 
$\psi(x_1,x_2)=\Psi(X) \,\psi(x)$, and
the eigen-energies the sum of the eigenvalues, $E=E_X+E_x$.
For the center of mass Hamiltonian Eq.(\ref{ehX}) we will consider
 the ground  state

\begin{equation}
\label{eXwf}
E_X=\frac{1}{2}\;\;\;;\;\;\;\Psi(X)\,=\,\frac{1}{\pi^{1/4}} e^{X^2/2} \,.
\end{equation}

This eigenfunction is symmetric, and the Hamiltonian 
Eq. (\ref{ehx}) is even in $x$, that means that the even (odd)
eigenfunctions of the Hamiltonian  Eq. (\ref{ehx}) correspond to
totally symmetric (antisymmetric) eigenfunctions under particle interchange.
The odd eigenfunctions of Hamiltonian Eq.~(\ref{ehx}) are the three-dimensional 
solutions written by Downing in Ref.~\cite{Downing2017} for zero angular 
momentum times $x$ (see Eq.~(\ref{etransasym})).

 Following~\cite{Downing2017}, in order to find the eigenfunctions of  the 
relative Hamiltonian Eq. (\ref{ehx}), we perform the transformations

\begin{equation}
\label{etranssym}
z=\left(\frac{x}{d}\right)^2\;\mbox{and}\; \psi(z)\,=\,e^{-d^2 z/2} y(z) 
\rightarrow \; \xi=-z \;\mbox{and} \; y(\xi)\,=\,(1-\xi) f(\xi) 
\end{equation}

\noindent for the symmetric eigenfunctions and
\begin{equation}
\label{etransasym}
z=\left(\frac{x}{d}\right)^2\;\mbox{and}\; \psi(z)\,=\,e^{-d^2 z/2} x\,y(z) 
\rightarrow \; \xi=-z \;\mbox{and} \; y(\xi)\,=\,(1-\xi) f(\xi) 
\end{equation}

\noindent for the antisymmetric ones. The function 
$f(\xi)$ fulfills the standard form of the Heun equation~\cite{Fiziev2010},

\begin{equation}
\label{eheun}
f''\,+\,\left(\alpha +\frac{\beta+1}{\xi}+\frac{\gamma+1}{\xi-1}\right) f'\,+\,
\left( \frac{\mu}{\xi}+\frac{\nu}{\xi-1}\right) f\,=\,0 \,,
\end{equation}

\noindent where the parameters are defined as

\begin{equation}
\label{eheunp}
\alpha\,=\,d^2\;;\;\beta\,=\,\mp \frac{1}{2} \;;\;\gamma\,=\,1 \;;\;
\mu\,=\,\frac{1}{4}\left(d^2 (1-k^2)+g-2\right)\;;\;\nu\,=\,d^2-\frac{g-2}{4} \,.
\end{equation}

\noindent and $k=2E_x$. The difference between
one-dimensional symmetric or antisymmetric functions  is given by the 
coefficient 
$\beta=-1/2$  and $\beta=1/2$, respectively.

 As usual \cite{Fiziev2010}, we define the parameters  
\begin{equation}
\label{eed}
\eta\,=\,\frac{1}{2}(\alpha-\beta-\gamma+\alpha \beta-\beta \gamma)-\mu\,=\,
\frac{1}{4} (d^2 k^2-g+2) \;\;\;;\;\;\;
\delta\,=\,\nu-\eta-\frac{1}{2}(\alpha+\beta+\gamma+\alpha \gamma+\beta \gamma)
\,=\,-\frac{d^2 k^2}{4} \,,
\end{equation}

\noindent and the Heun function is written as

\begin{equation}
\label{echf}
f(\xi)\,=\,\sum_{n=0}^\infty \, v_n(\alpha,\beta,\gamma,\delta,\eta)\,\xi^n \,,
\end{equation}

\noindent where the coefficients are given by the recurrence relation

\begin{equation}
\label{err}
A_n v_n\,=\,B_n v_{n-1}\,+\,C_n v_{n-2} \;\;;\;\;v_0=1\;;\;v_{-1}=0 \,
\end{equation}

\noindent where

\begin{subequations}
\begin{eqnarray}
A_n&=&1+\frac{\beta}{n}  \;, \label{ean}\\
B_n&=&1+\frac{-\alpha+\beta+\gamma-1}{n}+\frac{
\eta-(-\alpha+\beta+\gamma)/2-\alpha \beta/2+\beta \gamma/2}{n^2} \label{ebn}\,,\\
C_n&=&\frac{\alpha}{n^2}\left(\frac{\delta}{\alpha}+\frac{\beta+\gamma}{2}+n-1\right)
\label{ecn} \,.
\end{eqnarray}
\end{subequations}

We note that the parameters $\alpha,\, \gamma,\, \delta$ and $\eta$ and the 
recurrence
relations are those of the three dimensional bosonic case for 
zero angular moment \cite{Downing2017}.

The confluent Heun functions are not
square-integrable \cite{Heun1} and the series must be truncated in order to
obtain a polynomial of degree $N$ in Eq.(\ref{echf}), which implies, from Eq.~(\ref{err}), $v_{N+1}=v_{N+2}=0$. 
Therefore the condition for the eigen-energies is $C_{n=N+2}=0$ in (\ref{ecn}), which gives 
\begin{equation}
\label{eenkn}
k_N^2\,=\,2 E_x\,=\,4 N+6 \mp 1 \;\Rightarrow \;E_N=E_x+E_X\,=\,2 N+\frac{7\mp1}{2}\,,
\end{equation}

\noindent where the upper (lower) sign describes symmetric (antisymmetric) states.
It is important to note that $E_N$ does not depend on $g$ or $d$,
then the regularized Calogero model is solvable over a discrete set of isoenergetic 
curves, that must reduce for $d=0$ to the polynomial solutions of the Calogero model,
 which are defined by an index $p$ \cite{opos15}, 
 corresponding to the parametrization 

\begin{equation}
\label{edp}
g^{(N)}_p(d=0)\,=\,p (p-1)\;, p=2,3,4\ldots \;.
\end{equation}

\noindent  Note that $N$ and $p$ are not  quantum numbers, so we will obtain ground- and 
excited-state functions for different values of  $N$ and $p$.

On the other hand, we know that the complete spectrum of the Calogero model 
is $E_n= n+\sqrt{1+4 g}/2+3/2$ \cite{Calogero1969},  where  $n$ is the principal 
quantum 
number. Please note that, as the Calogero model is one-dimensional and the potential diverges for $x=0$, both 
symmetric and antisymmetric solutions have the same $n$, and hence the same energy $E_n$. 
However, \emph{polynomial} solutions for symmetric (antisymmetric) states are 
only know for even (odd) $p\in \mathbb{N}$.  That is, for the discrete set of
polynomial solutions we obtain

\begin{equation} 
\label{eeb}
E_{n;p}= n+p+1 \;;\;n=0,2,\ldots\;;\;p=2,3,4\ldots \;,
\end{equation}

\noindent  and since $E_{n;p}$ and $E_N$ belong to the same isoenergetic curve,
$n,p$ must satisfy the relation

\begin{equation}
\label{eeN}
E_{n;p}\,=\,E_N \Rightarrow  n+p=2 N+\frac{5\mp1}{2}\,.
\end{equation}


\section{Symmetric eigenfunctions with $N=0$ and $N=1$}\label{sec:symmetric}

The formal expression for the general solutions are quite involved, 
but it is useful, for better comprehension, to write down the analytical expressions for wave functions,
reduced density matrices and entropies for the cases $N=0$ and $N=1$.

\subsection{N=0 }

\begin{figure}
\begin{center}
\includegraphics[width=7.cm]{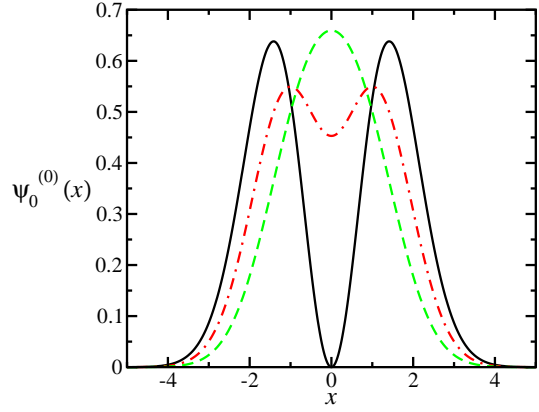}
\end{center}
\caption{  \label{fwfbN0} (Color online) Symmetric ground state wave function for $N=0$, $\psi_{0,2}^{(0)}(d)$, for three
different values of the cutoff length $d=0,1,2$ (full black, dash-dotted red and dashed green lines correspondingly).}
\end{figure}

In this case, Eq.(\ref{ecn}) gives  $k_N^2=5$, or $E_{N}=3$. 
The condition $v_1=0$ gives

\begin{equation}
\label{evb0}
g=2+4 d^2 \,,
\end{equation}

\noindent which gives $g(d=0)=2$ and so $p=2$. Hence we recast $g$ in 
Eq.~(\ref{evb0}) as $g^{(0)}_2(d)$. From
Eq.~(\ref{eeN}), we have $n=0$, that is, $N=0$ which corresponds to a ground state.
The wave function for the reduced coordinate is given by

\begin{equation}
\label{ewb0}
\psi_{0,2}^{(0)}(x)\,=\,\frac{2 (d^2+x^2)}{\pi^{1/4} 
\sqrt{3+4 d^2+4 d^4} } e^{-x^2/2} \,,
\end{equation}

\noindent where the subscripts and superscripts are chosen according 
to the prescription $\psi_{n,p}^{(N)}(x)$. It is interesting to note that $\psi_{0,2}^{(0)}(0)$ is a
minimum (maximum) for $d < (>) \sqrt{2}$. This phenomenon is shown  in 
Fig. \ref{fwfbN0}, where we plot the wave function in Eq.~(\ref{ewb0}) for $d=0, d=1$ and $d=2$.

The 1-RDM takes the form

\begin{equation}
\label{eq:erb0}
\rho_{0,2}^{(0)}(x,y)\,=\,\frac{ e^{-(x^2+y^2)/2}}{4\,\sqrt{\pi}\, 
(3+4 d^2+4 d^4) } (3+8 d^2+16 d^4+2 (1+4 d^2) (x^2+y^2)+
8 x y+4 x^2 y^2) \,.
\end{equation}

Using the orthonormal Hermite functions

\noindent 
\begin{equation}
\label{ehof}
\psi_k(x)\,=\,\frac{e^{-\frac{1}{2} x^2}\,H_k(x)}{\sqrt{2^k k! \pi^{1/2}}}\,,
\end{equation}

\noindent where $H_k(x)$ are the Hermite polynomials, the  
 $1$-RDM can be written as 

\begin{equation}
\label{erho2d}
\rho^{(0)}_{0,2}({x};{y})\,=\, \sum_{i,j=0}^2
\,\rho_{i,j}(d)\,\psi_i({x})\,\psi_j({y})\,.
\end{equation}

\noindent The $1$-RDM above can be cast in matrix form

\begin{equation}
\label{ern0}
\left[\rho_{0,2}^{(0)}\right](d)\,=\,\left( \begin{array}{ccc}
 \frac{3/8+d^2+d^4}{3/4+d^2+d^4}&0&
\frac{1+2 d^2}{4 \sqrt{2} (3/4+d^2+d^4)}\\
 0&\frac{1}{4 (3/4+d^2+d^4) }&0\\
\frac{1+2 d^2}{4 \sqrt{2} (3/4+d^2+d^4)}&0&
\frac{1}{8(3/4+d^2+d^4) }\end{array} \right) \,,
\end{equation}

\noindent and its eigenvalues can be exactly calculated and are given by

\begin{subequations}
\begin{eqnarray}
\lambda_\pm&=&\frac{2+4 d^2 +4 d^4 \pm
(1 + 2 d^2) \sqrt{ 3 + 4 d^2 +4  d^4}}{
2 (3 + 4 d^2 + 4 d^4)} \nonumber \\
\mbox{} && \label{eavN0B}\\
\lambda_o&=&\frac{1}{4 (3/4+d^2+d^4) }\label{eavN0Bi}\,.
\end{eqnarray}
\end{subequations}

\begin{figure}
\begin{center}
\includegraphics[width=7.cm]{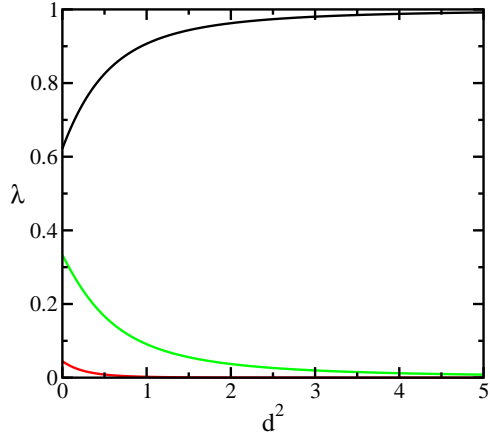}
\end{center}
\caption{  \label{favN0} (Color online) Eigenvalues of the $N=0$ reduced density 
matrix for symmetric solutions, $\rho^{(0)}_{0,2}(d)$ (Eq.~(\ref{eq:erb0})), as a function of the squared cutoff length $d^2$.}
\end{figure}

\noindent These eigenvalues  are showed in Fig.~\ref{favN0}.
In the important limit $d \rightarrow 0$ we obtain

\begin{equation}
\label{ern0c}
\left[\rho_{0,2}^{(0)}\right](0)\,=\,\left( \begin{array}{ccc}
 \frac{1}{2}&0&\frac{1}{3 \sqrt{2}}\\
 0&\frac{1}{3}&0\\
 \frac{1}{3 \sqrt{2}}&0&\frac{1}{6}\end{array} \right) \,,
\end{equation}

\noindent whose eigenvalues are

\begin{equation}
\label{eavn2}
\lambda_\pm\,=\,\frac{2\pm\sqrt{3}}{6}\;;\;\lambda_o\,=\,\frac{1}{3}\,,
\end{equation}

\noindent as reported in Ref.~\cite{opos15}. For 
$d\rightarrow \infty$, replacing $g_2^{(0)}(d)$ in Eq.~(\ref{eq:ehcal}), the interaction potential becomes a constant, and
the 1-RDM correspond to two non-interacting harmonic particles in
the ground state,

\begin{equation}
\label{ern0h}
\lim_{d\rightarrow \infty}\left[\rho_{0,2}^{(0)}\right](d)\,=\,\left( \begin{array}{ccc}
1&0&0\\
 0&0&0\\
 0&0&0\end{array} \right) \,.
\end{equation}

\subsection{N=1}

In this case $k_N^2=9$, which implies $E_{N}=5$.
The value for $v_1$ is

\begin{equation}
\label{evb1}
v_1(d,g)\,=\,1+4 d^2-\frac{g}{2} \,,
\end{equation}

and the condition $v_2=0$ gives

\begin{equation}
\label{egb1}
g_\pm\,=\,7+6 d^2 \pm\sqrt{25-12 d^2+4 d^4}\,.
\end{equation}

\noindent Performing the same analysis for the functions $g_{\pm}(d)$ as 
was done previously for $g_2^{(0)}(d)$, we get that 
$g_+(d=0)=12 \Rightarrow p=4 \Rightarrow g_+(d)=g^{(1)}_4(d)$, and similarly $g_-(d)=g^{(1)}_2(d)$. Then,
Eq.~(\ref{eeN}) gives two solutions for the energy, $E_{0,4}$, corresponding to
the ground state for $g=12$, and $E_{2,2}$ corresponding to the second excited
state for $g=2$. $g^{(1)}_4$ corresponds to the ground state, with the nodeless wave function

\begin{equation}
\label{ewfb1p}
\psi_{0,4}^{(1)}(x)\,=\,\frac{2 \sqrt{2} \left(2 d^4+d^2 (7-2 d^2+D) x^2+(5-2 d^2+D) x^4
\right) e^{-x^2/2}}{\pi^{1/4} \sqrt{525 (5+D)+
2 d^2 \left(60 +75 D+2 d^2 \left( 2 d^2 (8+6 d^2-D)-3 (8+D)\right)\right)}} \;.
\end{equation}

\noindent where $D=\sqrt{25+4 (d^2-3) d^2}$.
The limit cases of this wave function for $d\rightarrow 0$ and $d\rightarrow \infty$
are 

\begin{equation}
\label{ewfb1plc}
\left. \psi_{0,4}^{(1)}(x)\right|_{d=0}\,=\,\frac{4 x^4 e^{-x^2/2}}{\pi^{1/4} \sqrt{105}}
\;\;\;;\;\;\;
\left. \psi_{0,4}^{(1)}(x)\right|_{d\rightarrow \infty}\,=\,\frac{e^{-x^2/2}}{\pi^{1/4} } \;,
\end{equation}

\noindent the first one correspond to the Calogero ground state for $g=12$, the 
second one to the ground state of two non interacting particles in an harmonic potential. The 1-RDM
is a $5 \times 5$ matrix.

Taking $g^{(1)}_2$ we obtain the second excited state, whose wave function is

\begin{equation}
\label{ewfb1m}
\psi_{2,2}^{(1)}(x)\,=\,\frac{2 \sqrt{2} \left(2 d^4+d^2 (-7+2 d^2+D) x^2+(5-2 d^2-D) x^4
\right) e^{-x^2/2}}{\pi^{1/4} \sqrt{-525 (-5+D)+
2 d^2 \left(60 -75 D+2 d^2 \left( 2 d^2 (8+6 d^2+D)+3 (-8+D)\right)\right)}} \;,
\end{equation}

\noindent which has two nodes. The limit of this wave function for $d\rightarrow 0$ is

\begin{equation}
\label{ewfb1mlc}
\left. \psi_{2,2}^{(1)}(x)\right|_{d=0}\,=\,\frac{\sqrt{2}}{\sqrt{15} \pi^{1/4}}
 x^2 (2 x^2-5)  e^{-x^2/2} \,,
\end{equation}

\noindent which  corresponds to the second excited state for the Calogero model with
$g=2$, and for the limit  $d\rightarrow \infty$  is

\begin{equation}
\label{ewfb1mlc_in}
\left. \psi_{2,2}^{(1)}(x)\right|_{d\rightarrow \infty}\,=\,
\frac{H_2(x) e^{-x^2/2}}{2 \sqrt{2} \pi^{1/4} } \;.
\end{equation}

\noindent The complete wave function is given by

\begin{equation}
\label{ewfb1mlcC}
\left. \psi(x_1,x_2)\right|_{d\rightarrow \infty}\,=\,\Psi(X)\,\left. \psi_{2,2}^{(1)}(x)\right|_{d\rightarrow \infty}\,=\,
\frac{1}{2} \left(\psi_2(x_1)\,\psi_0(x_2)+\psi_0(x_1)\,\psi_2(x_2)\right)
-\frac{1}{\sqrt{2}}
\psi_1(x_1)\,\psi_1(x_2)\;,
\end{equation}

\noindent note that these three terms are the only products of Hermite 
functions that give the energy $E=5$ and they share equal probability 
amongst even and odd eigenfunctions.

\section{Antisymmetric eigenfunctions with N=0}\label{sec:anti-symmetric}

%
%
%
%
%

The energy in this case
is $k_N^2=7$ or $E_{N}=4$, and the condition $v_1=0$ gives

\begin{equation}
\label{evf0}
g^{(0)}_3=6+4 d^2 \,,
\end{equation}

\noindent then, from  Eq.(\ref{eeN}),
it correspond to the antisymmetric  ground state  $E_{0,3}$
where the reduced density matrix of the Calogero model is exactly solved for 
fermions~\cite{opos15}. The reduced wave function is

\begin{equation}
\label{ewf0}
\psi_{0,3}^{(0)}(x)\,=\,\frac{2 \sqrt{2} x (d^2+x^2)}{\pi^{1/4}
\sqrt{15+12 d^2+4 d^4} } e^{-x^2/2} \,.
\end{equation}

The 1-RDM takes the form

\begin{eqnarray}
\label{eq:erf0}
\rho_{0,3}^{(0)}(x,y)\,=\,\frac{e^{-(x^2+y^2)/2}}{\pi^{1/2} 8\left(15+12 d^2+4 d^4\right) } & &\left( 15+24 d^2+16 d^4+(54+48 d^2+32 d^4) xy \right.\nonumber\\
& + & 6 (3+4 d^2) (x^2+y^2)+4 (3+4 d^2) x y (x^2+y^2) \nonumber\\
& + & \left.36 x^2 y^2+8 x^3 y^3 \right)\,.
\end{eqnarray}

\begin{figure}
\begin{center}
\includegraphics[width=7.cm]{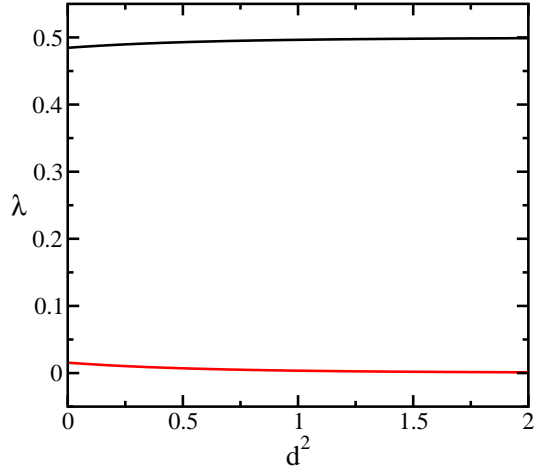}
\end{center}
\caption{  \label{favFN0} (Color online) Eigenvalues of the $N=0$ reduced density  
matrix for antisymmetric solutions, $\rho^{(0)}_{0,3}$ (Eq.(\ref{eq:erf0})), as a function of the squared cutoff length $d^2$. Note that both eigenvalues 
are doubly degenerate.}
\end{figure}

The complete matrix in the Hermite basis set, Eq. (\ref{ehof}), is
\begin{equation}
\label{emrf0}
\left[\rho^{(0)}_{0,3}\right](d)\,=\,\left( \begin{array}{cccc}
 \frac{21+24 d^2+8 d^4}{4 (15+12 d^2+4 d^4)}&0&\frac{3\sqrt{2} (3+2 d^2)}{4 (15+12 d^2+4 d^4)}&0\\
 0&\frac{27+24 d^2 +8 d^4}{4 (15+12 d^2+4 d^4)}&0& \frac{\sqrt{6} (3+2 d^2)}{4 (15+12 d^2+4 d^4)}\\
\frac{3\sqrt{2} (3+2 d^2)}{4 (15+12 d^2+4 d^4)}&0&\frac{9}{4 (15+12 d^2+4 d^4)} &0\\
0& \frac{\sqrt{6} (3+2 d^2)}{4 (15+12 d^2+4 d^4)}&0& \frac{3}{4 (15+12 d^2+4 d^4)}
\end{array} \right) \,,
\end{equation}

\noindent and its eigenvalues, shown in Fig.~\ref{favFN0},  are

\begin{equation}
\lambda_\pm\,=\,\frac{1}{4}\left[1\pm\frac{\sqrt{2 (99+4 d^2 (3+d^2) (15+2 d^2 
(3 + d^2))}}{(15+2 d^2 (3 + d^2))} \right]\;, \label{eevfn0p}\\
\end{equation}

\noindent both with multiplicity 2. In the limit $d\rightarrow 0$ we obtain the 
expressions
 reported for the two-fermion Calogero model \cite{opos15}

\begin{equation}
\label{ern0fd0}
\left[\rho^{(0)}_{0,3}\right](0)\,=\,\left( \begin{array}{cccc}
 \frac{7}{20}&0&\frac{3}{10 \sqrt{2}}&0\\
 0&\frac{9}{20}&0& \frac{1}{10}\sqrt{\frac{3}{2}}\\
 \frac{3}{10 \sqrt{2}}&0&\frac{3}{20} &0\\
0& \frac{1}{10}\sqrt{\frac{3}{2}}&0& \frac{1}{20}
\end{array} \right) \,,
\end{equation}

\noindent whose eigenvalues are

\begin{equation}
\label{eavn2f}
\lambda_\pm\,=\,\frac{5\pm\sqrt{22}}{20}\,.
\end{equation}

For the limit $d\rightarrow \infty$ we get

\begin{equation}
\label{ern0fdi}
\lim_{d\rightarrow \infty}\left[\rho^{(0)}_{0,3}\right](d)\,=\,\left( \begin{array}{cccc}
 \frac{1}{2}&0&0&0\\
 0&\frac{1}{2}&0& 0\\
0&0&0 &0\\
0& 0&0&0
\end{array} \right) \,,
\end{equation}

\section{The R\'enyi and von Neumann entropies} 
\label{sec:renyi}
So far, we have been involved with the quasi-exactly solvable aspect of the 
regularized Calogero model. This Section is devoted to analyze both the behavior of 
the entanglement entropies for the exact polynomial solutions and for arbitrary pairs of the pair $(g,d)$. 

\begin{figure}
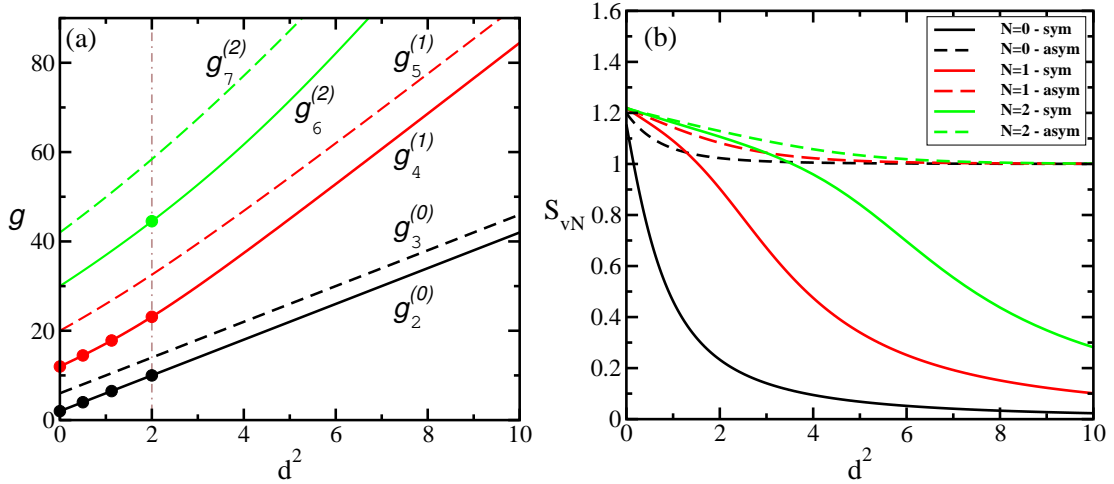

\begin{center}
\includegraphics[width=7.cm]{fig4a.eps}
\includegraphics[width=7.5cm]{fig4b.eps}
\end{center}
\caption{  \label{fig:svn-gd} 
(Color online) (a) Isoenergetic curves $g^{(N)}_p(d)$ where an exact polynomial solution of the 
one-dimensional regularized Calogero model is known. From bottom to top the solid (dashed) lines correspond to
symmetric (antisymmetric) ground-state wave functions for $N=0,1,2$. 
The brown dash-dotted line corresponds to $d^2=2$ used to compute the approximate eigenvalues of Fig.~\ref{fig:eigenvalues-rho} (a) and the circles on this line are those shown at the bottom of the same figure and pinpoint the $g$ values for which a number of eigenvalues of the approximate $1$-RDM become null. Correspondingly, dots on the $g^{(0)}_2(d^2)$ and $g^{(1)}_4(d^2)$ symmetric curves correspond to those shown at the bottom of Fig.~\ref{fig:drift}.  (b) Exact von Neumann entropy of the ground-state wave functions obtained for
the isoenergetic curves $g^{(N)}_p(d)$. The same color code is used in both panels. }
\end{figure}

For a given density matrix $\rho$ with an entanglement 
spectrum $\{\lambda_i\}$, its spectral decomposition is

\begin{equation}
 \rho= \sum_i \lambda_i \left|{\phi_i}\right\rangle\left\langle\phi_i\right|,
\end{equation}

\noindent where the $\left|{\phi_i}\right\rangle$'s are the eigenvectors or 
natural orbitals of $\rho$. The eigenvalues $\{\lambda_i\}$ are also known as 
natural occupation numbers.

The R\'enyi entropy of $\rho$ is defined as

\begin{equation}\label{eq:renyi-entropy}
S^{a} (\rho) = \frac{1}{1-a} \log_2 \mbox{Tr} \rho^{a} ,
\end{equation}
where $a>0$ is a constant (here we use $a$ instead of the more common $\alpha$ 
parameter to avoid possible conflicts with the parameter of Eq.~(\ref{eheunp})). 
Besides, it is well known that
\begin{equation}\label{eq:vonneumann-entropy}
 \lim_{a \rightarrow 1}  S^{a} (\rho) = S_{vN}(\rho) = - \mbox{Tr}(\rho \log_2 
\rho) ,
\end{equation}
where $S_{vN}(\rho)$ is the von Neumann entropy. In some cases, the 
mono-parametric family of R\'enyi entropies shed more light over the 
peculiarities of the entanglement spectrum, {\em i.e.} the spectrum of the 
$\rho$ under study, because of its ability to weight differently the 
eigenvalues of $\rho$ by changing the value of $a$. This is made clear 
by looking at the expressions of  both entropies, Eqs.~(\ref{eq:renyi-entropy}) 
and (\ref{eq:vonneumann-entropy}), in terms of the eigenvalues of $\rho$

\begin{equation}\label{eq:entropies-eigenvalues}
S^a(\rho) = \frac{1}{1-a} \log_2 \left(\sum_i \lambda_i^a \right), \qquad 
S_{vN} = - \sum_i \lambda_i \log_2 \lambda_i .
\end{equation}

Let us start by inspecting the isoenergetic $g^{(N)}_p(d)$ curves
in the $(g,d^2)$ plane. Figure~\ref{fig:svn-gd}(b) show these curves for all
the ground states $n=0$. Note that the curves hit the ordinate axis at the $g$
values where an exact polynomial solution of the Calogero model can be
found~\cite{opos15}. The von Neumann entropies (vNE) along each isoenergetic
curve are presented in Fig.~\ref{fig:svn-gd}(a). The vNE for the
symmetric case goes to zero for large values of the cutoff length parameter
indicating a single natural orbital population. Conversely, in 
the antisymmetric case the vNE converges to a limiting value of one because
the antisymmetrization prevents such single natural orbital population. 
Note that the vNE at $d=0$ is not the same for all curves and whether symmetric or antisymmetric configurations have larger vNE depends upon
the particular $g^{(N)}_{p}(0)$ value as shown in~\cite{opos15} for the 
Calogero model.

\begin{figure}
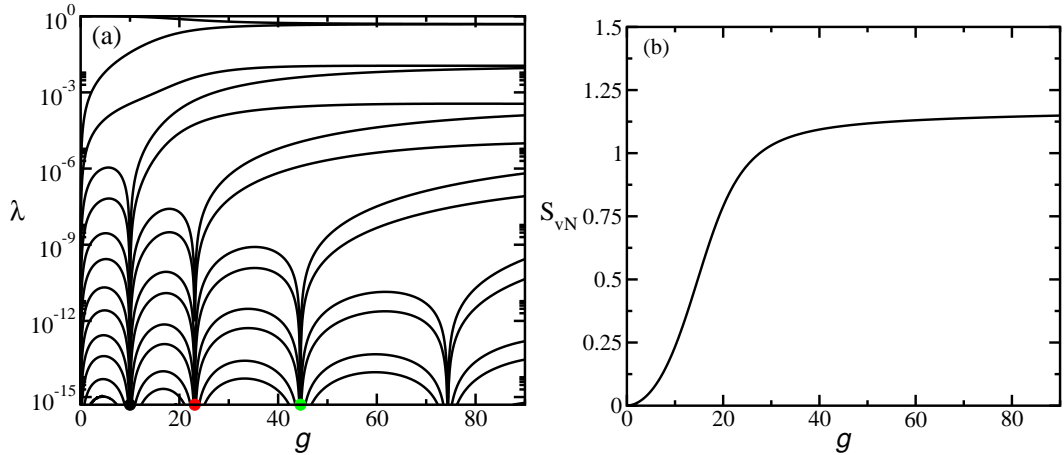

\begin{center}
\includegraphics[width=7cm]{fig5a.eps}
\includegraphics[width=7.cm]{fig5b.eps}
\end{center}
\caption{  \label{fig:eigenvalues-rho} (Color online) (a) Approximate eigenvalues of the $1$-RDM as a function of $g$.
The $1$-RDM was obtained from the variational wave function approximation to the symmetric 
ground state wave function for $d^2=2$. The coloured dots at the bottom indicate the $g$ values
for which a number of eigenvalues become null, and are also indicated in Fig.~\ref{fig:svn-gd}(a). 
(b) Approximate von Neumann entropy corresponding to the eigenvalues shown in panel (a).}
\end{figure}

We turn now to the study of the $1$-RDM eigenvalues and vNE for arbitrary values
of the parameters in the $(d^2,g)$-plane.  
Figure~\ref{fig:eigenvalues-rho}(a) shows the typical behavior of the largest 
eigenvalues of the $1$-RDM for $d\neq 0$ as a function of $g$. The eigenvalues 
were calculated using a high precision variational method with a symmetrical 
Hermite-DVR basis set function~\cite{Manthe2017,Beck2000}, so the eigenvalues 
corresponds to a symmetric problem. The method to obtain the eigenvalues of the 1-RDM from the 
approximate two-particle variational wave function has been discussed 
elsewhere \cite{Osenda2007,Osenda2008,Giesbertz2013}. There is a discrete set 
of $g$ values where a number of eigenvalues become null. From these points the 
eigenvalues are increasing functions of $g$, but only the two largest do not 
become null again. The rest reach some maximum and then become decreasing 
functions of $g$.   
After they go to zero, the two major eigenvalues do not become null 
again and so on.

The values of $g$ where a number of eigenvalues become null for a given 
fixed value of $d$ can be plotted in the plane $(d^2,g)$. 
Figure~\ref{fig:svn-gd}(a) shows these values for $d^2=2$ as filled dots and 
they match those shown at the bottom of Fig.~\ref{fig:eigenvalues-rho}(a). 
It is clear that the values of $g$ where a number of 
eigenvalues become null coincide with those found in the previous Sections, 
since the curves shown in Fig.~\ref{fig:svn-gd}(a) correspond to the 
analytical equations found for the lowest eigenvalues corresponding to 
symmetric and antisymmetric functions, see  Eqs.~(\ref{evb0}),(\ref{egb1}) y 
(\ref{evf0}). Let us remember that the curves are isoenergetic, since the 
corresponding eigenvalue does not depend on $g$ or $d$ over the curve. Besides, 
the number of non-zero natural occupation numbers over each curve is always the same, and 
coincides with the number found in Sections~\ref{sec:symmetric} and 
\ref{sec:anti-symmetric}. From bottom to top in Fig.~\ref{fig:svn-gd}(a) the number is equal to three, 
five, and so on, for the symmetric eigenvalues. The same can be said for the 
eigenvalues corresponding to the  antisymmetric eigenfunctions.

\begin{figure}
\begin{center}
\includegraphics[width=7.cm]{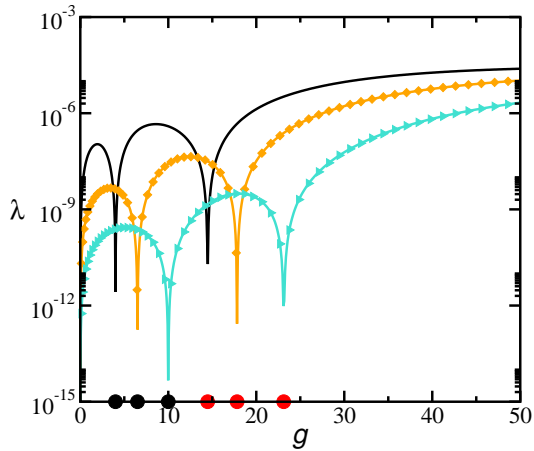}
\end{center}
\caption{  \label{fig:drift} (Color online) 7th eigenvalue obtained from the $1$-RDM constructed from the 
variational approximation to the symmetric ground state wave function. The curves are for three values of the
squared cutoff length $d^2=0.5, 1.125, 2$ (black, red and green correspondingly). 
The dots on the bottom indicate the interaction values $g$ for which
the eigenvalue shows a behavior that resembles that of an almost vanishing eigenvalue. 
Those points are the same shown in Fig.~\ref{fig:svn-gd}(b) over the isoenergetic curves $g^{(0)}_2(d^2)$ and $g^{(1)}_4(d^2)$ for symmetric solutions.}
\end{figure}

Since the isoenergetic curves $g^{(N)}_p = g^{(N)}_p(d^2)$ are increasing functions of 
$d^2$, it is clear that the values of $g$ where a number of eigenvalues become 
null are also increasing functions of $d$. This can be appreciated in 
Figure~\ref{fig:drift} where the seventh eigenvalue of the 1-RDM of 
the symmetric two-particle wave function is shown for several values of $d$. 
The sixth and seventh eigenvalues are the largest eigenvalues that have only 
two zeros. If $\lambda_i$ is the i-th eigenvalue of the 1-RDM, and $g_n^i$ 
is the $n$-th value of $g$ such that $\lambda_i(g_n^i)=0$, then 
$g^i_n<g^i_{n+1}$ and $g^i_{n}(d_1) < g^i_{n}(d_2)$, $\forall  \;d_1 < 
d_2$.

The R\'enyi entropies also provide a way to identify models where the number of 
eigenvalues of the RDM that are different from zero alternates between infinity 
and a finite value. This is the case of the Calogero model for which it has 
been shown that the entanglement spectrum has a numerable infinite 
number of non-zero elements for open sets of the interaction parameter, besides 
these open sets are separated between them by a discrete set of values of the 
interaction parameter, $g_n$, where the number of non-zero eigenvalues of the 
entanglement spectrum is finite \cite{Garagiola2016}.  As has been shown 
above, for the regularized Calogero model the set of values of the parameter 
where the entanglement spectrum is finite depends on the actual value of $d$, 
so $g_n=g_n(d)$.

The eigenvalues  of the 1-RDM, at a fixed value of $d$, are analytical 
functions of $g$, and this can be exploited to assume a concrete analytical 
expression for the eigenvalues. As a consequence, explicit
expressions for the Renyi entropies and its derivatives can be written. We  develop 
here the case for symmetric two-particle wave function (the anti-symmetric case 
is similar), where the 1-RDM has only $2n+1$ non-zero 
eigenvalues at $g=g_n$, in the following the dependency with $d$ is dropped to 
keep the notation as simple as possible.

The following results will only rely on the analyticity of the eigenvalues 
around isolated points in the parameter space where the spectrum is finite. 
Assuming that

\begin{equation}
\label{eq:aee}
\lambda_i(g)\sim \left\{ \begin{array}{lll}
\lambda_i(g_n) +    \lambda_i^{(1)} (g-g_n) \; &\mbox{ if } \;i\le2 n+1\\
\mbox{} & \hspace{4cm} \mbox{ for } g \rightarrow g_n \\
 \lambda_i^{(2)}(g-g_n)^{2 k_{i,n}}  
\;\;\;\;\;\;\;&\mbox{ if } i>2 n+1\,,
\end{array} \right.
\end{equation}

\noindent where $\lambda_i^{(1)}, \lambda_i^{(2)}$ are constants, and 
$k_{i,n}\ge 1$ is an integer. Eq. (\ref{eq:entropies-eigenvalues}) can be 
written as

\begin{eqnarray}
\label{eq:Re-b}
S^a(\nu)&=&
\frac{1}{1-a} \, \log_2{\left(\sum_{i=1}^{2n+1} \lambda_i^a(g)  +
{\sum_{i=2n+2}^\infty  \lambda_i^a(g) } \right)} \nonumber \\
\mbox{}&=& 
\frac{1}{1-a} \, \left(\log_2{\left(\sum\limits_{i=1}^{2n+1} 
\lambda_i^a(g)\right) } +
\log_2{\left(1+\frac{\sum\limits_{i=2n+2}^\infty  \lambda_i^a(g) }{
\sum\limits_{i=1}^{2n+1} \lambda_i^a(g)} \right)}\right)   \\
\mbox{}& \underset{g\rightarrow g_n}{\sim}&
\frac{1}{1-a} \, \left(\log_2{\left(\sum\limits_{i=1}^{2n+1} 
\lambda_i^a(g)\right) } +
\frac{\sum\limits_{i=2n+2}^\infty  \lambda_i^a(g) }{ \ln{2}\,
\sum\limits_{i=1}^{2n+1} \lambda_i^a(g)} \right) \,=\,
S_n^a(g)+s_n^a(g)\,. \nonumber
\end{eqnarray}

\noindent The last line of the equation above is the definition of the 
quantities $S_n^a(g)$ and $s_n^a(g)$. So, it is clear that 
$S_n^a(g_n)=S^a(g_n)$, and 
$s_n^a(g_n)=0$. Then, the derivative of the R\'enyi entropy at 
$g=g_n$ can be obtained as

\begin{eqnarray}
\label{eq:dsdnu-asym}
\left.\frac{\partial S^a(g) }{\partial g}\right|_{g=g_n}&=&
\left.\frac{\partial S_n^a(g) }{\partial g}\right|_{g=g_n} +  
\nonumber \\
\mbox{}&&\frac{a}{\ln{2} \, (1-a)} \, \left(
 \frac{\sum\limits_{i=2n+2}^\infty \lambda_i^{a-1}(g)
\partial_g \lambda_i(g)}{\sum\limits_{i=1}^{2n+1} \lambda_i^a(g)}
-\frac{\sum\limits_{i=2n+2}^\infty  \lambda_i^a(g) 
\sum\limits_{i=1}^{2n+1} \lambda_i^{a-1}(g)
\partial_g \lambda_i(g)}{\left( \sum\limits_{i=1}^{2n+1} 
\lambda_i^a(g)\right) ^2} 
\right)_{g=g_n} \,. 
\end{eqnarray}

\noindent The first  term in Eq. (\ref{eq:dsdnu-asym}) is a well-defined 
constant
and the third one is zero. As a result of this,  the derivative is dominated 
by the second term. Using the analytic expansion of the eigenvalues, 
Eq. (\ref{eq:aee}), and assuming that 
$k_m$  is the minimum value of $k_{i,n}$, the 
leading asymptotic behavior of $s_n^a$ is

\begin{equation}
\label{eq:sa}
s_n^a(g) \underset{g\rightarrow g_n}{\sim}  
C_n\,((g-g_n)^{2 k_m})^{a} \,=\, C_n \,
|g-g_n|^{\chi k_m}\,,
\end{equation}

\noindent where $\chi=2 a$, which implies that

\begin{equation}
\label{eq:dsdnua}
\frac{\partial s_n^a(g) }{\partial g}\, \underset{g\rightarrow 
g_n}{\sim}
\chi k_m C_n\,|g-g_n|^{\chi k_m-1} \,sign(g-g_n)\,.
\end{equation}

\noindent Collecting the results of the last few equations, the derivative of 
the R\'enyi entropy can be expressed as

\begin{equation}
\label{eq:dsdnun}
\left.\frac{\partial S^a(g) }{\partial g}\right|_{g=g_n}\,=\,
\left\{ \begin{array}{lrl} \left. \begin{array}{lll}
-sign(C_n ) \times \infty  & \mbox{for } &g \rightarrow g_n^- \\
sign(C_n) \times \infty  & \mbox{for } &g \rightarrow g_n^+
 \end{array}  \right\}  
& \mbox{if }\; \chi k_m <1  \\
 \left.\begin{array}{lrl}
\partial_g S_n^a(g_n)\,- C & \mbox{for } &g \rightarrow g_n^-\\
\partial_g S_n^a(g_n)\,+ C  & \mbox{for } &g \rightarrow g_n^+ 
\end{array}  \right\}
  & \mbox{if }\;\chi k_m =1  \\
\partial_g S_n^a(g_n)   & \mbox{if }\; \chi k_m\ge 1  \,.
  \end{array} \right.
\end{equation}

\begin{figure}
\begin{center}
\includegraphics[width=7.cm]{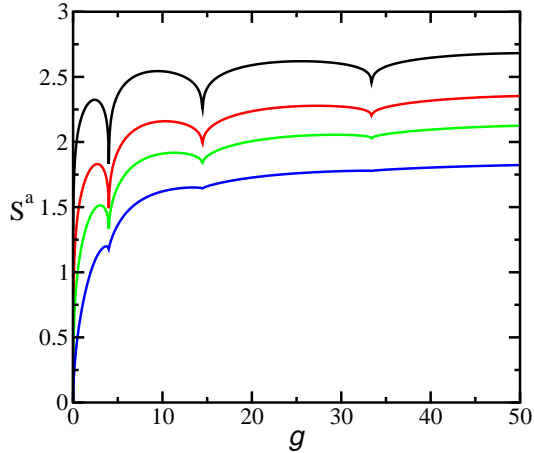}
\end{center}
\caption{  \label{fig:renyies} (Color online) R\'enyi entropies as a function of the interaction parameter $g$
for a cutoff length $d^2=0.5$. The parameter values $a=0.1,0.15,0.2,0.3$ 
(black, red, green and blue line correspondingly) were chosen. The kinks are located at $g$ values for which there is an exact polynomial
expansion of the $1$-RDM, which renders a finite Hilbert space and hence many eigenvalues vanish.
As shown in the text, the kinks are a consequence of the analyticity of the vanishing eigenvalues around those $g$ values.}
\end{figure}

\noindent Even tough the derivative of $S^a$ is continuous for 
$\chi \ge 1$, it is straightforward to see from the eigenvalue asymptotics, 
Eq. (\ref{eq:aee}), that the second derivative diverges for $1<  \chi k_m < 2$, 
but it is analytical for $ \chi k_m=2$, \textit{i.e} the kink at $\chi k_m=1$ 
is smoothed until it disappears at $ \chi k_m=2$.

Figure~\ref{fig:renyies} shows the behavior of $S^a$  for different values of 
the parameter $a$, as a function of $g$, while $d=$ is kept fixed. The kinks at 
fixed values of $g$ can be easily appreciated, as well as their softening for 
increasing values of $a$ as predicted by Eq.~(\ref{eq:dsdnun}), observe 
that the bottom curve corresponds to the largest value of $a$ depicted while 
the upper curve corresponds to the lower one. Keeping $d$ fixed  assures that 
the points $g_m$ where only a  number of eigenvalues are non-zero are also kept 
fixed and, as a consequence, the kinks in the curves calculated for different 
values of $a$ are located at the same abscissas. 

\section{discussion}
\label{sec:discussion}

Few particle models with  exact and finite reduced density matrices are 
even more scarce than models with exact solutions, making them valuable 
examples to test numerical methods to obtain the spectrum of the Hamiltonian and 
the entanglement spectrum. Recently, there has been a number of works dealing 
with the properties of the entanglement spectrum, or natural occupation numbers, 
in particular the phenomenon of {\em pinning}. Much of the understanding has 
been obtained analyzing systems of coupled harmonic oscillators (or Moshinsky 
model), because they are amenable of a complete analytical treatment. It is 
feasible that the study of quasi-solvable models helps the efforts made to 
understand the pinning phenomenon and other related issues.

Quasi-exactly solvable models have wave functions that are polynomial functions on 
the inter-particle distance so, at least for those that do not depend on any 
angular variable but the ones on the inter-particle radius, they  should 
also possess exact and finite reduced density matrices. This last problem is open for 
three dimensional problems with non-trivial angular momentum. 

For the model analyzed in this work, the quasi-exactly solvable character is 
intertwined with the fact that the Calogero model has exact solutions that can 
be expressed as polynomials in the interparticle distance. So, when we take 
the limit $d\rightarrow 0$ {\em over} the isoenergetic lines we are able to 
recover all the quantities corresponding to the Calogero model. Then, it is 
natural to wonder if a given quasi-solvable model has always, in some limit, 
an unknown exactly solvable relative. 

The R\'enyi entropy shows, again, that it is a capable tool to identify systems 
with exact and finite RDM. Nevertheless, to improve its usability it is necessary to 
determine if a set of very small eigenvalues are 
effectively zero or not. To accomplish this it is necessary to identify if, for 
example, performing a finite size analysis of the numerical eigenvalues at the parameter 
where the system has an exact and finite RDM the behaviour is (quite) different 
from the behavior where there is not such a RDM. It is clear that for models 
with wave functions with only a polynomial dependency on the 
inter-particle distance it is possible to choose a finite basis set that 
expands the Hilbert space where the wave function to be analyzed is contained 
exactly, resulting in an exact RDM. In this case, the RDM derived from the 
finite basis contains all the information required to produce a finite number 
of non-zero eigenvalues and a number of exactly zero ones. Conversely, when the 
finite basis set used to analyze a given wave function does not contain the 
exact wave function under consideration there will be a number of 
eigenvalues that should be zero in the limit of an infinite basis set, but for 
a finite basis they are not and a numerical criterion is in order. Work around 
these lines is in progress.


\end{document}